\documentclass[preprint2]{aastex}

\newcommand{\nh}{N_{\rm H}}
\newcommand{\wfe}{W_{\rm b}}

\newcommand{\g}{$\gamma$}

\newcommand{\sax}{{\it Beppo\-SAX}}
\newcommand{\source}{NGC\,4151}

\newcommand{\asca}{{\it ASCA}}
\newcommand{\gro}{{\it CGRO}}

\begin{document}

\title{NGC 4151: AN INTRINSICALLY AVERAGE SEYFERT 1}

\author{Andrzej A. Zdziarski,\altaffilmark{1}
Karen M. Leighly,\altaffilmark{2}
Masaru Matsuoka,\altaffilmark{3}
Massimo Cappi,\altaffilmark{4}
and Tatehiro Mihara\altaffilmark{5}
}
\altaffiltext{1}{Centrum Astronomiczne im.\ M. Kopernika, Bartycka 18,
00-716 Warszawa, Poland; aaz@camk.edu.pl}
\altaffiltext{2}{Department of Physics and Astronomy, The University
of Oklahoma, 440 W.\ Brooks St., Norman, OK 73019, USA; leighly@ou.edu}

\altaffiltext{3}{NASDA-SURP 2-1-1, Sengen, Tsukuba, Ibaraki
305-8505, Japan}

\altaffiltext{4}{Istituto Te.S.R.E, CNR, Via Gobetti 101, 40129 Bologna, Italy}

\altaffiltext{5}{RIKEN, 2-1 Hirosawa Wako, Saitama, 351-0198, Japan}

\begin{abstract} We present a detailed analysis of a long, 100 ksec, observation
of \source\ by \asca, contemporaneous with an observation by the \gro/OSSE. We
fit the data with physical models including an Fe K line and Compton reflection
both relativistically broadened and coupled according to theoretical results.
The model also includes a narrow Fe K component, which is emitted by both an
extended plasma region and the X-ray absorber. Our study of the absorber shows
strong evidence for the presence of more than one partial-covering cloud in the
line of sight. Taking these points into account, our best intrinsic model
includes a Comptonization continuum with an X-ray slope of $\Gamma\simeq 1.9$
from a thermal plasma with an electron $kT \sim 70$ keV and a disk line with an
equivalent width of $\sim 70$ eV. The broadening of the line and reflection
indicate their origin from innermost parts of an accretion disk. Our results
indicate that \source\ has an intrinsic X-ray spectrum and variability
properties similar to those of average Seyfert 1s. \end{abstract}


\section {Introduction}
\label{intro}

\source, a Seyfert 1 galaxy at $z=0.0033$, has the $\sim 2$--10 keV photon index
of its power-law component reported from observations by several different
satellites to be in the range of $0.3\la \Gamma\la 1.8$ (Yaqoob \& Warwick 1991;
Yaqoob et al.\ 1993; Wang et al.\ 1999, hereafter W99; Ogle et al.\ 2000; Yang
et al.\ 2001; Piro et al.\ 2002). (Note that the lowest of these values,
0.3--0.4, have been measured by {\it Chandra}.) The corresponding average
Seyfert 1 photon index is $\Gamma= 1.95$ with an intrinsic standard deviation of
only 0.15 (Nandra \& Pounds 1994), so the lower values in the above range are
quite unusual.  Since \source\ is the most often studied Seyfert (as it is the
brightest one in the hard X-rays), it is of great interest to determine whether
it is an archetype of its class or very far from it, as would be indicated by
the spectral index varying up to $11\sigma$ from the average.

Reliable determination of the spectral indices of power law indices in accreting
black holes is of major theoretical importance. Our present understanding of the
power law component in both Seyferts and black hole binaries associates its
origin with the process of thermal Comptonization (e.g., Poutanen 1998;
Zdziarski 2000). However, this process yields in general $\Gamma>1$ (e.g.,
Sunyaev \& Titarchuk 1980). If, on the other hand, Comptonization were
non-thermal, $\Gamma\geq 1.5$ (including cooling and pair production, e.g.,
Lightman \& Zdziarski 1987). Very hard local continua can be obtained from
Comptonization in a highly relativistic plasma ($kT\ga 500$ keV) of a small
optical depth (e.g., Haardt 1993), but this model for \source\ is then in
conflict with the measurement of its plasma temperature of $kT<100$ keV (Johnson
et al.\ 1997, hereafter J97; Zdziarski et al.\ 2000). If the measurement of
$\Gamma\sim 0.3$--0.4 in \source\ (Ogle et al.\ 2000; Yang et al.\ 2001) is
indeed confirmed, it would call for a new physical explanation. Furthermore,
Yang et al.\ (2001) found no contribution of the scattered component to the soft
excess in \source.  However, this result was based on their value of
$\Gamma=0.3$ obtained assuming a single uniform absorber.  On the other hand,
the fitted value of $\Gamma$ strongly depends on the model of the complex
absorber in this source, as noted by Weaver et al.\ (1994, hereafter W94).

Also, the Fe K$\alpha$ line has been found to be very broad and strong in
\asca\/ observations during hard states of \source\ (Yaqoob et al.\ 1995; W99).
A small $\Gamma$ during those observations would constitute an interesting
counterexample to the positive correlation between the strength and relativistic
broadening of Fe K reprocessing features and $\Gamma$ observed in Seyferts
(Zdziarski et al.\ 1999; Lubi\'nski \& Zdziarski 2001), which is both expected
theoretically and confirmed to hold in black-hole binaries (Gilfanov et al.\
2000).

On the other hand, monitoring by the OSSE detector aboard \gro\/ shows the shape
of its 50--300 keV spectrum to be weakly variable and very similar to the
corresponding average spectra of Seyferts (J97; Zdziarski et al.\ 2000). In
particular, the model of thermal Comptonization and Compton reflection (standard
Seyfert continuum components) applied to the average OSSE spectrum of \source\
yields the 2--10 keV index of $\Gamma\sim 1.9$ (J97), i.e., very close to the
measured average of Seyfert 1s.

It is quite possible that the unusual properties of \source\ noted above arise
because the X-ray spectrum of \source\ is usually strongly absorbed with a
characteristic column density of $\nh\sim 10^{23}$ cm$^{-2}$, which makes
measuring the intrinsic X-ray index difficult (as already noted by W94). For
instance, the observation with the historical weakest absorption (Yaqoob et al.\
1993), $\nh\sim 10^{22}$ cm$^{-2}$, has also the highest historical 2--10 keV
flux, a soft $\Gamma\simeq 1.8$ and a low upper limit on the Fe K line (see also
Zdziarski et al.\ 1996, hereafter Z96). This may indicate that an apparent
correlation between the 2--10 keV flux and $\Gamma$ (e.g., Yaqoob et al.\ 1993)
is to a large degree due to strong absorption, which both suppresses the emitted
X-ray flux and hardens the spectrum (although Yaqoob \& Warwick 1991 have
rejected that possibility based on their treatment of the absorption).

Z96 have suggested previously that the low values of $\Gamma$ and/or the very
strong Fe K line may be artifacts of insufficient account of complex absorption
and neglect of Compton reflection. However, as noted by W99, the data used by
Z96 were either of a low signal-to-noise ratio or limited energy resolution.
Subsequently, W99 found the Fe K line to be very strong and broad and
$\Gamma\simeq 1.3$ in a long, 100 ksec, \asca\/ observation in 1995, and
commented on the unprecedented high signal to noise of that observation.

Here, we re-examine the above issues by reanalyzing the 1995 \asca\/
observation.  Our treatment is different from that of W99, who considered only
the 1--10 keV SIS data, as we analyze the full \asca\/ spectrum together with a
contemporaneous OSSE observation and employ relatively more sophisticated
physical models. Preliminary results of our study of the \asca\/ spectrum were
given in Leighly et al.\ (1997, 1998). In our present best model, the intrinsic
$\Gamma\simeq 1.9$, equal to that obtained from modelling the soft \g-rays. We
conclude that previous reports of very low values of $\Gamma$ in \source\ are
likely to be due to simplified treatments of the X-ray absorption and neglect of
Compton reflection. We find the broad line to have a moderate equivalent width
of $\sim 10^2$ eV, which is average for Seyfert 1s with that $\Gamma$
(Lubi\'nski \& Zdziarski 2001). Also, we find that the energy dependence of the
variability amplitude implies a major role of complex variable absorption, in
agreement with the results of the spectral fitting.

\section{Data Analysis}

\begin{figure}[t!]
\epsscale{1.0}
\plotone{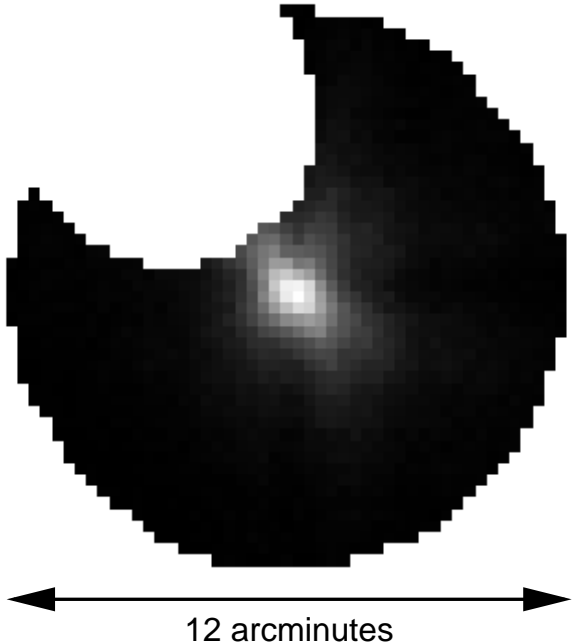}
\caption{The part of the \asca\/ GIS2 image of the NGC 4151 used in the present
reduction. The region around the nearby BL Lac 1E1207.9+3945 has been removed.
\label{f:image}
}
\end{figure}

\source\ was observed by \asca\/ from 1995 May 10 02:49:43 till 1995 May 12
13:50:36 (UT, for the SIS) with the net exposure of $\simeq 10^5$ s for each of
the detectors. SIS 1-CCD faint mode was used and standard data reduction was
applied. Care was taken to exclude photons from the BL Lac object 1E1207.9+3945
(which lies $5\arcmin$ from NGC 4151), as shown in Figure \ref{f:image}.  We
have rebinned the original channels to 1/4 of the FWHM resolution of the SIS and
GIS detectors, and assumed a 1.5\% systematic error, which was estimated from
residual discrepancies among the 4 \asca\/ detectors. We note that this
observation was made only 2 years after the \asca\/ launch, and thus was only
slightly affected by the time-dependent degradation of the SIS resolution.

We fit the data using XSPEC (Arnaud 1996). We assume elemental abundances from
Anders \& Ebihara (1982) and the interstellar Galactic absorbing column of
$2.17\times 10^{20}$ cm$^{-2}$ (Murphy et al.\ 1996). The single-parameter
uncertainties below correspond to 90\% confidence ($\Delta \chi^2=2.7$).

\section{The Average Spectrum}

\subsection{Simple phenomenological models}
\label{simple}

We start with a model similar to one of those used by W94, consisting of a power
law, assumed here to be e-folded with $E_{\rm c}=400$ keV, absorbed uniformly
with a column density, $N_0$, and partially with a column $N_1$ covering a
fraction $f_1$ (the so-called dual absorber). The soft component is modeled by
the sum of a fraction, $f_{\rm sc}$, of the power law scattering towards the
observer (suffering only the Galactic absorption) and thermal bremsstrahlung.
The model also includes a Gaussian Fe K$\alpha$ line absorbed only by the
Galactic $\nh$, which is fitted to be narrow with the width of $\sigma_{\rm
n}\simeq 0.1$ keV and the peak energy of $E_{\rm n} \simeq 6.3$ keV.

\begin{figure*}[t!]
\epsscale{2.2}
\plottwo{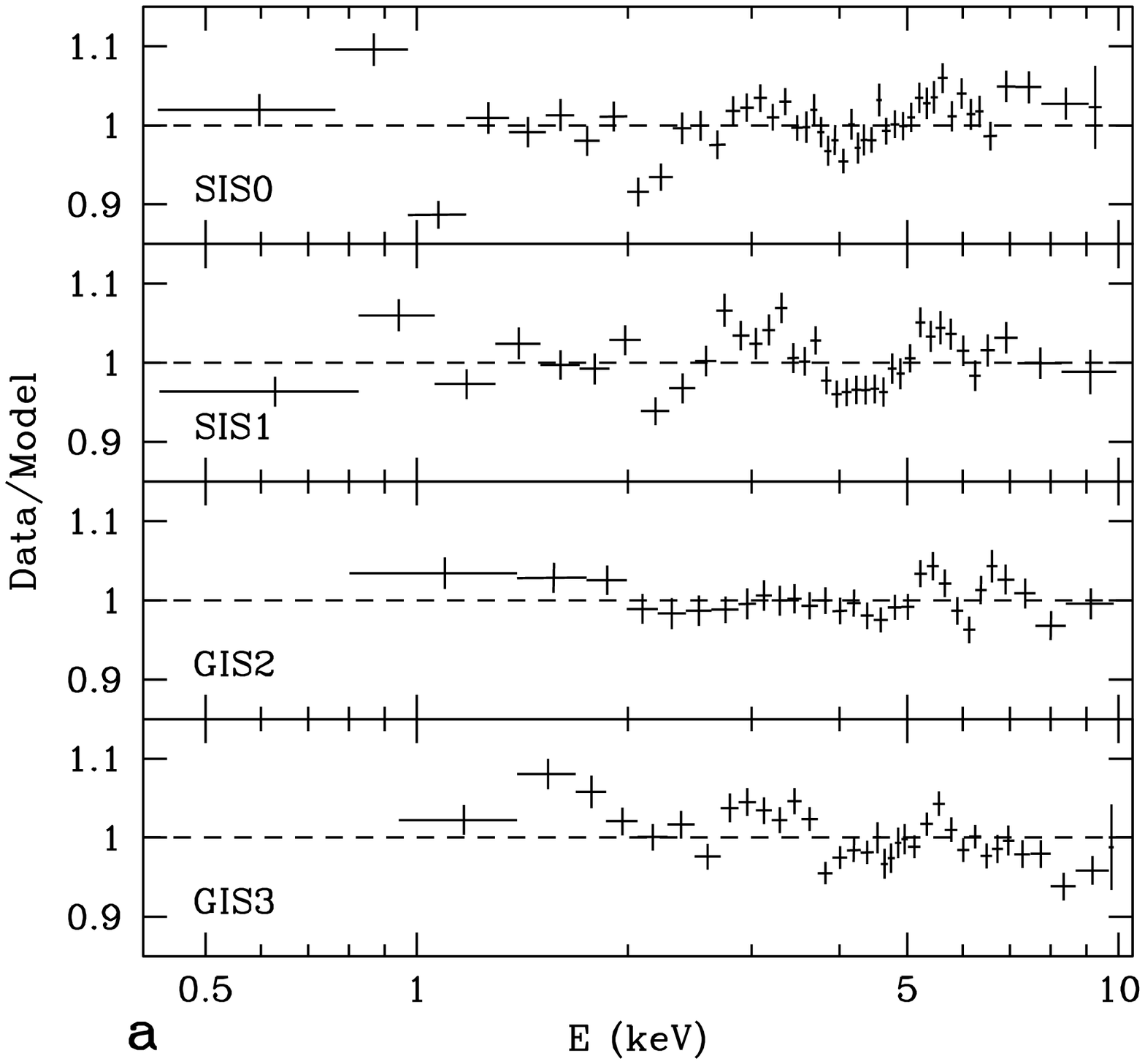}{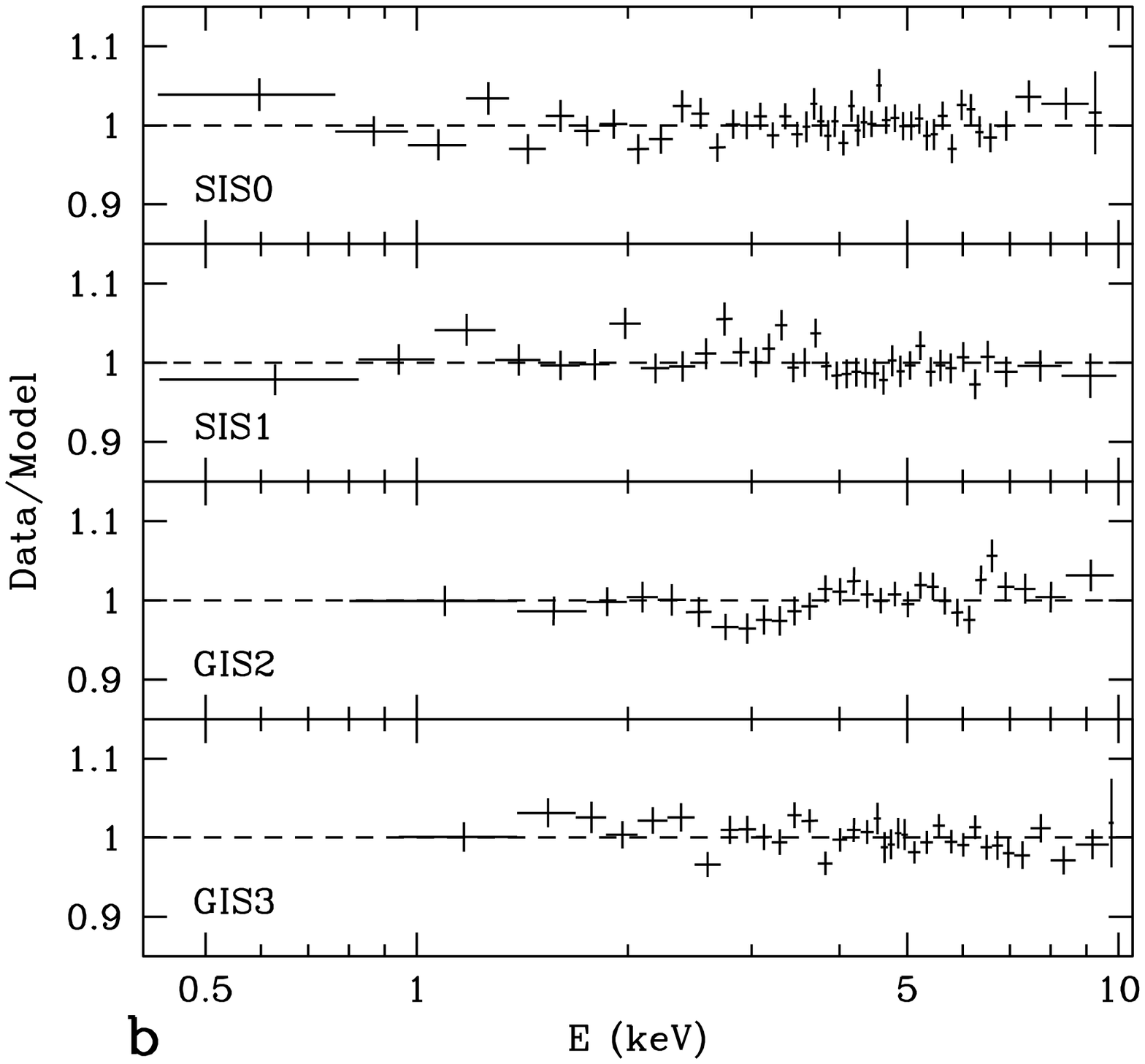}
\caption{(a) The residuals to the model with a power law, dual absorber, a
Gaussian line and bremsstrahlung. (b) The residual for our best model for the
\asca\/ data, obtained by adding a disk line with associated reflection,
ionization of the full screen, and an additional partially-covering cloud.
(The assumed systematic error of 1.5\% is not shown.)
\label{f:residuals}
}
\end{figure*}

Such a model provided a good fit ($\chi^2/\nu\sim 1$) to the two $\sim 20$ ksec
net-exposure data taken in 1993 (W94). However, it provides a very poor model to
the present data, $\chi^2/\nu=1016/641$, and it yields strong localized
residuals, including one around $\sim 5$--6.5 keV, as illustrated in Figure
\ref{f:residuals}a. A likely reason for this difference is the much better
statistics of the present data, which has a longer exposure by a factor of $>5$.

In particular, including (in addition to the narrow Gaussian) a broad Gaussian
feature (absorbed as the power law) with the peak energy at $E_{\rm b}=
5.9^{+0.1}_{-0.2}$ keV, the width of $\sigma_{\rm b}= 0.9^{+0.2}_{-0.1}$ keV,
and equivalent width of $\wfe=390$ eV, reduces $\chi^2$ by 141. Furthermore, the
data show line-like residuals at 0.89 keV, and addition of a narrow line further
reduces $\chi^2$ by 123.  This feature probably corresponds to the O {\sc viii}
radiative recombination continuum and Ne {\sc ix} emission that originates in
the X-ray emitting narrow-line region, and is seen clearly in the {\it
Chandra\/} spectrum (Ogle et al.\ 2000). Also, we find that the SIS continuum is
systematically harder than the GIS one. This is likely to be due to the residual
inaccuracy of the SIS calibration for the present very hard X-ray spectrum, and
is recognized here because of the very good statistics. Addition of a small
power law correction to the SIS model, $\Delta\Gamma$, reduces $\chi^2$ by 41
(chance probability is $2\times 10^{-9}$; hereafter, we use the F-test to
estimate significance of adding model components). Hereafter, we keep the soft
line energy and $\Delta\Gamma$ fixed at 0.89 keV and $+0.047$, respectively.  We
note that W99, who analyzed the SIS data at energies $\geq 1$ keV only, have not
considered the complexities of the soft spectrum present in the data.

This model, with $\chi^2/\nu=711/637$, which is within $\sim 2\sigma$ of the
mean of the $\chi^2$ distribution, is formally acceptable but it has a number of
problems. First, at the bremsstrahlung temperature of 0.31 keV, strong line
emission is expected. However, using the {\tt mekal} model in XSPEC (Mewe et
al.\ 1985) actually worsens the fit by $\Delta \chi^2 =+16$. Second, the origin
of the very strong and broad Gaussian line is unclear. It is indicative of very
strong Compton reflection (e.g., \.Zycki \& Czerny 1994), which has not been
included in the model.  The continuum spectral index is $\Gamma=
1.39^{+0.05}_{-0.05}$, and Compton reflection is negligible in other Seyferts
with a such a hard spectrum (Zdziarski et al.\ 1999). Thus, this model can be
considered only an empirical (rather than physical) description of the spectrum.

\subsection{Fe K line profile and Compton reflection}
\label{reflection}

To make further progress, we have tried to develop a physical description of the
spectrum by taking into account the following issues.  The formation of Fe
fluorescence lines proceeds through photoionization, so that the number of
photons removed beyond the iron K-edge times the fluorescence yield equals the
number of line photons. Broad Fe K$\alpha$ lines are most likely formed by
reflection from a cold medium (Nandra \& Pounds 1994). In this case, basic
physics demands that the line flux should be tied to the relative strength of
Compton reflection, $\Omega/2\pi$ (with the coefficient of proportionality
dependent on the ionization state; George \& Fabian 1991; \.Zycki \& Czerny
1994; \.Zycki et al.\ 1998). Furthermore, if a line is broadened and shifted
kinematically and by gravity, the same broadening will occur to the reflection
component, in particular its K edge. Then, models with broad lines originating
from illumination of an accretion disk but either without reflection or with
static reflection are not physically self-consistent.

We note that there is evidence that in some Seyferts with observed
broad Fe K lines the simple model with irradiation of an accretion
disk extending close to the minimum stable orbit may not be applicable
(e.g., Chiang et al.\ 2000).  Possibly, some of the red wings observed
in Seyferts are due to a continuum complexity not understood at the
present time rather than due to an irradiated accretion disk. On the
other hand, the irradiated disk model, including the associated
edge/reflection component, works well in some other cases, e.g., in
the archetypical Seyfert 1 object IC 4329A (Done et al.\ 2000). In any
case, a failure of the irradiated disk model cannot be considered a
justification for fitting X-ray spectra with the disk line only and
without the associated reflection component. Instead, new physical
models for the red wings should be looked for.

The photoionization process preceding the Fe K$\alpha$ fluorescence should be
neglected only when an edge/reflection component is so weak that it is
observationally negligible. This is the case, in particular, for lines emitted
by a Thomson-thin plasma (e.g., Makishima 1986). Such a narrow line is already
included in our model, and it is due to both extended emission and the emission
of the absorbing clouds. However, such lines are highly unlikely to be strongly
Doppler/gravity broadened as an optically-thin accretion flow is necessarily
very hot close to the central black hole (e.g., Narayan \& Yi 1995; Abramowicz
et al.\ 1995).

Accordingly, we replace the power-law continuum in the last model in \S
\ref{simple} by the sum of a power law and Compton reflection normalized in such
way that the line equivalent width with respect to the Compton-reflected
component alone at 6.4 keV is $\simeq 1.3$ keV (George \& Fabian 1991).
This coefficient corresponds to a neutral or moderately ionized
reflection, and we check this assumption {\it a posteriori} below. The
reflection component (Magdziarz \& Zdziarski 1995) is then Gaussian-broadened
and redshifted in the same way as the line. The origin of the Gaussian
broadening may be the chaotic motion of clouds in the vicinity of the black
hole. This model yields a much softer intrinsic continuum, with $\Gamma=
1.87^{+0.06}_{-0.10}$, but very strong reflection, $\Omega/2\pi =
3.4^{+0.7}_{-1.0}$ ($\wfe\simeq 390$ eV), at $\chi^2/\nu=715/637$. We assumed
here a face-on inclination; for a larger inclination, the fitted $\Omega/2\pi$
becomes even larger.

As noted in \S \ref{simple}, Compton reflection as strong as this is unusual for
Seyferts. Also, it is likely that accretion proceeds via a disk rather than
clouds, and thus we consider the above model rather implausible. Therefore we
replace the broad Gaussian line by a line from an accretion disk in the
Schwarzschild metric with the rest energy of 6.4 keV and the emissivity $\propto
r^{-\beta}$, extending from an outer radius, $r_{\rm out}$ (assumed here to
equal $10^3$) to an inner one, $r_{\rm in}\geq 6$, where $r$ is in the units of
$GM/c^2$ (Fabian et al.\ 1989). The strength of the line is coupled to that of
Compton reflection in the same way as above. The reflection spectrum is again
relativistically broadened in the same way as the line ({\tt refsch} in XSPEC).
Since we include reflection, we constrain the inclination to $i\geq 15\degr$,
which corresponds to the range at which the reflection Green's functions of
Magdziarz \& Zdziarski (1995) are valid.

\begin{figure*}[t!]
\epsscale{2.2}
\plottwo{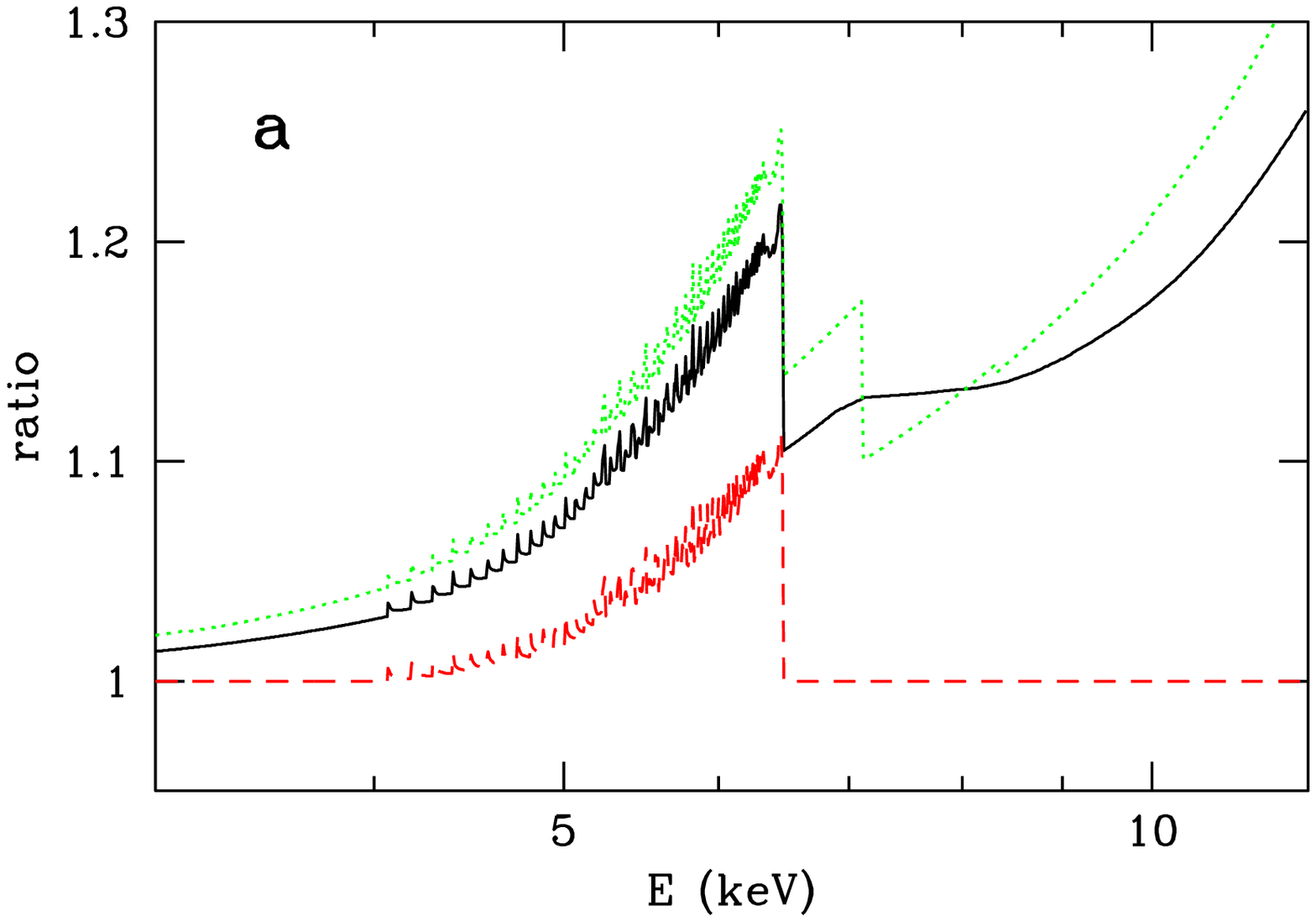}{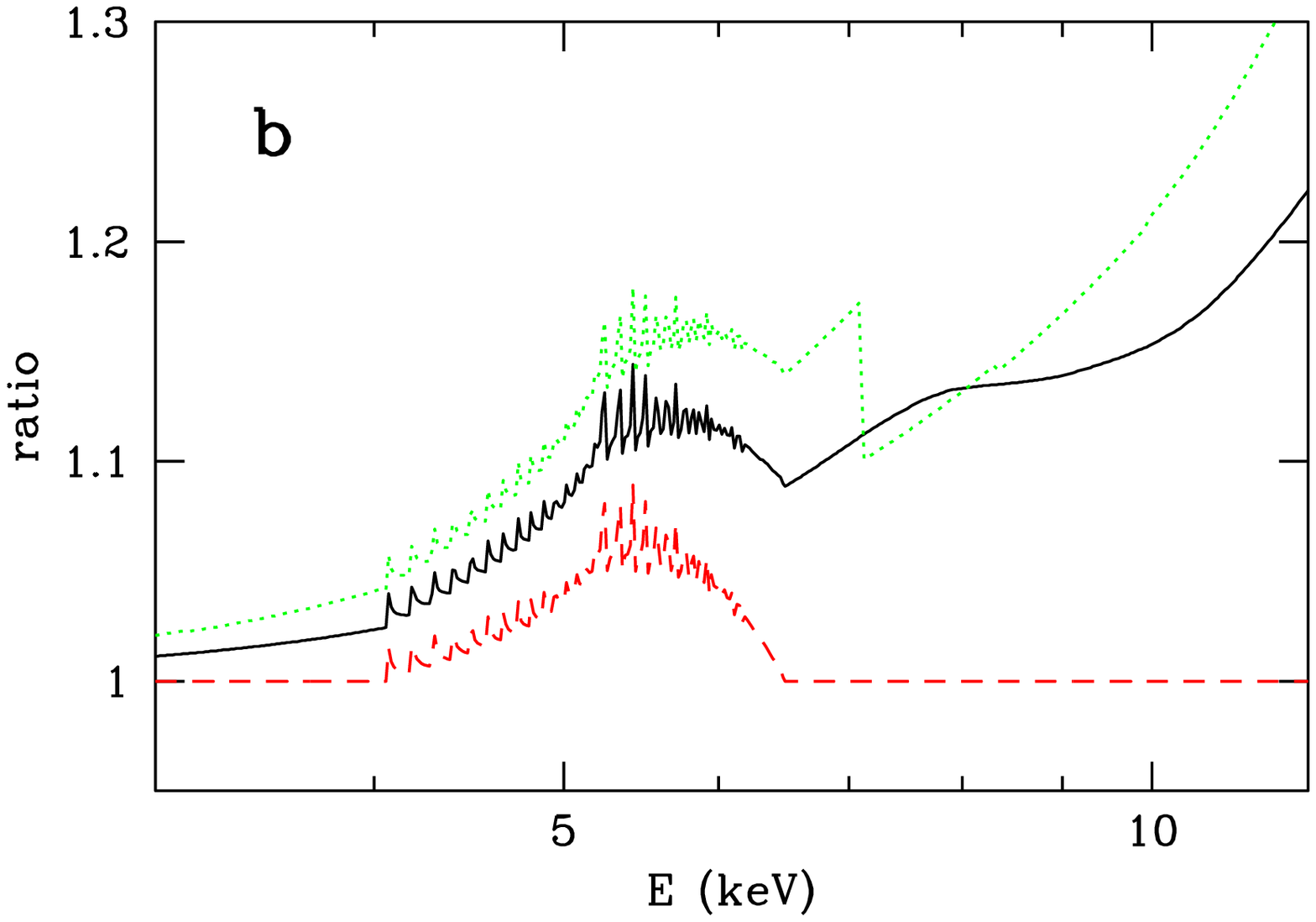}
\caption{The effect of Compton reflection on the Fe K$\alpha$ disk
line profile as illustrated by the ratio of the total spectrum to the
incident power law with $\Gamma=1.7$. The dashed curve shows the
case with reflection neglected, the dotted curves show
the line profile together with the associated reflection but
neglecting its own relativistic smearing, and the solid curves give
the correct spectrum in which both the line and the reflection
components are relativistically smeared in the same way. The
parameters are $r_{\rm in}=6$, $r_{\rm out}= 10^3$, $i=15\degr$, $\Omega/2\pi=1$
and (a) $\beta=3$ and (b) $\beta=4.5$. The wiggles in the line profile are
numerical noise in the XSPEC implementation of the disk line in Schwarzschild
metric ({\tt diskline}, Fabian et al.\ 1989). We note that this numerical noise
will be averaged out in data from \asca\/ given its resolution, but it prevents
the {\tt diskline} model from using in fits of high resolution data, e.g., those
from {\it Chandra}.
\label{f:line}
}
\end{figure*}

The importance of including Compton reflection in models of reprocessing
features is illustrated in Figure \ref{f:line} for two cases with strong
relativistic broadening in an accretion disk. We see that the effect is rather
strong, and the reflection component is by no means negligible. Also, taking
into account the relativistic broadening of the reflection component is crucial.
Roughly, the main effect of including relativistically-smeared reflection is an
overall hardening of the total continuum, but without appearance of any distinct
Fe K edge. The corresponding effect on the fitted parameters will be an increase
of the fitted $\Gamma$ of the incident spectrum.

We also include the K$\beta$ component (at the 7.06 keV rest energy and with the
flux of 0.12 of that of K$\alpha$) for both the disk and the narrow lines. Also,
we use the optically-thin plasma emission with the standard abundances (Mewe et
al.\ 1985) instead of bremsstrahlung. This model yields $\chi^2/\nu=701/636$,
$\Gamma= 1.63^{+0.04}_{-0.07}$, $\Omega/2\pi =1.34^{+0.17}_{-0.29}$ ($\wfe =
160$ eV).  The fit strongly prefers an inclination of the disk close to face-on,
$i=15^{+4}\degr$, and emission of the reprocessed component concentrated close
to the minimum stable orbit, $r_{\rm in}=6.2^{+1.1}_{-0.2}$,
$\beta=4.0^{+1.4}_{-0.4}$.

We note here that the quoted equivalent widths are given by XSPEC with respect
to the continuum at the peak of the line. E.g., for models with $\beta\simeq
4.5$, $i=15\degr$, $r_{\rm in}=6$, this peak is at $\sim 5.5$ keV (Fig.\
\ref{f:line}b), at which the continuum photon flux is typically by $\sim 30\%$
higher than that at 6.4 keV. Thus, for the same photon flux of the line, the
measured equivalent width will be correspondingly lower. This effect has to be
taken into account when making comparison with theoretical results giving the
line strength expected theoretically, which usually are expressed as the
equivalent width of the line in the local rest frame (e.g., George \& Fabian
1991; \.Zycki \& Czerny 1994).

\subsection{Complex absorption}
\label{absorption}

The complexity of absorption in \source\ has long been appreciated (e.g.\ W94).
Among others, Warwick et al.\ (1995) and George et al.\ (1998) have considered
absorption by a partially ionized medium. However, W99 have found that this
model provides a bad fit to the present data. On the other hand, ionization of
the completely-covering component in the dual absorber model has been found to
be required by Piro et al.\ (2002) for \sax\/ data. We confirm their conclusion
for our data, obtaining a reduction of $\chi^2$ by $20$ (the chance probability
of $2\times 10^{-5}$), a low ionization parameter of $\xi\simeq 0.6$ erg
s$^{-1}$ cm, but only small changes in the other fitted parameters. Here,
$\xi\equiv 4 \pi F_{\rm ion}/n$, where $F_{\rm ion}$ is the ionizing flux, $n$
is the reflector density, and the temperature of the reflecting medium is
assumed to be $10^5$ K (Done et al.\ 1992).

Next, we consider possible modifications of the partial covering component.
Specifically, since there is no reason {\it a priori} that there should be a
single partially-covering cloud in the line of sight, we investigate the
possibility of multiple clouds in two ways. First, we replace the partial
covering component by a Poisson distribution of clouds in the line of sight,
first proposed by Yaqoob \& Warwick (1991). However, this yields only a small
increase of $\Gamma$ and almost no change in $\chi^2$; thus, it is not required
statistically. Furthermore, the Poisson distribution requires an individual
event to have small probability, and this distribution would not be appropriate
if there are only a few clouds covering a substantial fraction of the source.
Thus, we consider the case of just two partially covering clouds (with the
column $N_2$ covering a fraction $f_2$ for the second one).  This leads to our
overall best model of the \asca\/ data, with $\chi^2/\nu = 655/633$. The
probability that the fit improvement with respect to the previous model (the
dual absorber with a weakly ionized full screen, see above) is by chance is
$4\times 10^{-6}$. This model yields $i= 17^{+3}_{-2}\degr$, $r_{\rm
in}=6.3^{+1.0}_{-0.3}$, $\beta= 5^{+2}_{-2}$, and a much softer continuum,
$\Gamma=1.87^{+0.07}_{-0.10}$, $\Omega/ 2\pi=1.20^{+0.54}_{-0.37}$ corresponding
to $\wfe=110$ eV at the best fit. Interestingly, these $\Gamma$ and $\Omega$ are
close to the Seyfert average (Zdziarski et al.\ 1999). The residuals for this
model are shown in Figure \ref{f:residuals}b.

\subsection{The broad-band 0.4--400 keV spectrum}
\label{broad}

As noted in \S 1, the \asca\/ observation was nearly contemporaneous with an
observation by OSSE, made 1995 Apr.\ 25--May 9. J97 has found that the
variability of \source\ in soft \g-rays is rather slow and modest. Thus, we fit
the OSSE spectrum (at a free relative normalization) together with the \asca\/
spectra.  We find that the last model in \S \ref{absorption} above predicts
quite well the amplitude (within 10\%) of the OSSE spectrum. The model
parameters are almost identical as for the model of the \asca\/ data alone
(e.g., $\wfe=120$ eV) except that now the e-folding energy is fitted.  The model
P in Table 1 shows the parameters (here, the standard values of $\beta=3$ and
$r_{\rm in}=6$ have been assumed). For this model, we also study the allowed
ranges of some parameters we otherwise keep fixed. We find the relative Fe
abundance of $0.9^{+0.2}_{-0.3}$, and, importantly, the reflecting medium to be
at most weakly ionized, with $\xi_{\rm refl}=1_{-1}^{+30}$ erg s$^{-1}$ cm,
which range corresponds to the dominance of Fe {\sc i--xii}, where effects of
reflector's ionization are negligible. In accordance with this result,
increasing the disk line energy to 6.7 keV (Fe {\sc xxv--xxvi}) leads to $\Delta
\chi^2=+18$, where we made conservative assumptions of a free $i$ and $\Omega$
independent of $\wfe$.

\begin{figure}[t!]
\epsscale{1.0}
\plotone{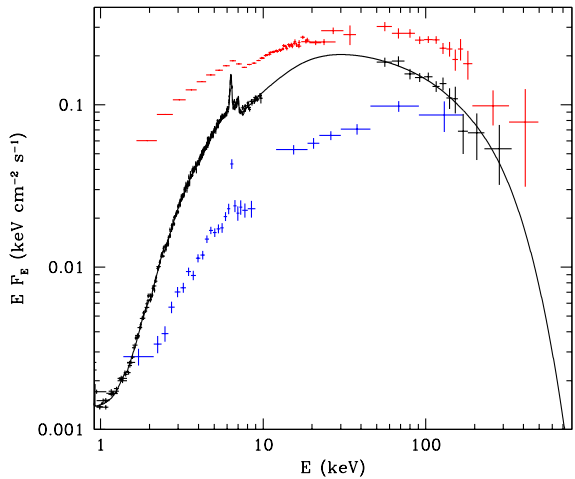}
\caption{The \asca-OSSE spectrum (black, middle), compared to the
historical range (blue to red) of X-ray and soft \g-ray fluxes (highest and
lowest X-rays from {\it Ginga\/} and {\it EXOSAT}, respectively, Z96; highest
and lowest soft \g-rays from OSSE, J97, and \sax/PDS, P.-O.\ Petrucci, private
communication, respectively). The solid curve gives the model with thermal
Comptonization (\S 3.3).
\label{f:compare}
}
\end{figure}

\begin{figure}[t!]
\epsscale{1.0}
\plotone{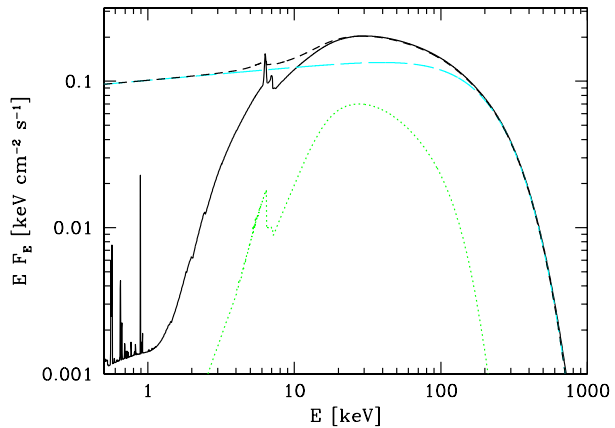}
\caption{The components of the model with thermal Comptonization (solid
curve). The short dashes show the intrinsic spectrum of the model with thermal
Comptonization before absorption. The long dashes and dots give the
Comptonization continuum and the reflection component together with the broad
Fe K lines, respectively.
\label{f:model}
}
\end{figure}

Next, we investigate a more physical model for the high energy spectrum.  The
high-energy cutoff of the OSSE data is strongly indicative of thermal Compton
scattering, and the average OSSE spectrum of \source\ can be well modelled by
it (J97). An e-folded power law is only a rough approximation to spectra from
that physical process. Thus, we now fit the spectrum with an isotropic
Comptonization model ({\tt compps} at
ftp://ftp.astro.su.se/pub/juri/XSPEC/COMPPS) of Poutanen \& Svensson (1996).
The parameters of the thermal Compton continuum are the plasma temperature,
$kT$, and the Compton parameter, $y\equiv 4\tau kT/m_{\rm e} c^2$, where $\tau$
is the plasma Thomson optical depth. The assumption that the seed photons are
blackbody at a temperature of 5 eV affects the resulting spectrum only very
weakly. We find the absorber and the soft component parameters are virtually
unchanged with respect to the previous model, but the reflection and disk line
(with $\wfe = 70$ eV) are somewhat weaker, see Table 1, model C.  The
normalization of the OSSE model with respect to that of SIS0 is close to unity
(1.08), and the 2--10 keV index of the Comptonization component alone is
$\Gamma=1.89$. The spectrum is shown in Figure \ref{f:compare}, together with
the range of X-ray and soft \g-ray spectra historically observed from \source.
Figure \ref{f:model} shows the main components of the model.

For this model, we also consider the effect of allowing for variable $\beta$
and $r_{\rm in}$. We find that adding these two free parameters results only in
a statististically insignificant reduction of $\chi^2$, by $-3$. At this best
fit, $\beta=3.9$, $r_{\rm in}= 6.3$, $\wfe\simeq 60$ eV.

We also note that we find a bad fit of the thermal Compton model when we use
the standard dual absorber model for the absorption ($\chi^2/\nu= 759/661$).
Also, that model requires that the normalization of the OSSE spectrum is up by
a factor of 2 with respect to that of the \asca\/ spectra. Thus, inclusion of
the OSSE data provides further very strong support for the presence of complex
absorption (with two partial-covering components). This result  is in agreement
with the OSSE spectra from \source\ being consistent with thermal
Comptonization with $\Gamma \simeq 1.9$ in X-rays (J97; Zdziarski et al.\
2000).

The total intrinsic model luminosity is $L_{\rm bol}= 3.9\times 10^{43}$ erg
s$^{-1}$, assuming isotropy and a distance of 13 Mpc. The emission measure of
the optically-thin plasma is then $1.0\times 10^{63}$ cm$^{-3}$, although we
stress that we have approximated its emission as collisionally-ionized only
(plus the 0.89-keV line) whereas Ogle et al.\ (2000) find evidence for both
collisional ionization and photoionization. Our final models have the best-fit
scattered fraction of $f_{\rm sc} \simeq 1.5\%$ (see Table 1), which agrees with
theoretical expectations of unified AGN models (Antonucci 1993). On the other
hand, the corresponding model with the standard dual absorber has a higher
value of $f_{\rm sc}\simeq 3.5\%$.

\subsection{Comparison with W99}
\label{w99}

W99 have studied a relatively small subset of the data considered here, namely
the SIS data in the 1--10 keV range. Their model consists of the dual absorber,
a power law component, Fe K disk and narrow lines from a neutral medium with
both K$\alpha$ and $K\beta$ components, and an unabsorbed (scattered) power law.
That last component is sufficient to model the soft excess in the considered
energy range as most of its complexities appear below 1 keV (see \S
\ref{simple}), but we note it yields a rather high $f_{\rm sc}\simeq 5\%$. W99
neglected reflection of the incident power law from the disk as they claimed
there is no statistical evidence for its presence. We note, however, that this
makes their model only a phenomenological description of the data, since their
model assumes irradiation of a nearly-neutral, optically-thick, disk by the
power-law source, and under these conditions a part of the irradiating photons
{\it will\/} be Compton backscattered (see \S \ref{reflection}).

\begin{figure}[t!]
\epsscale{1.0}
\plotone{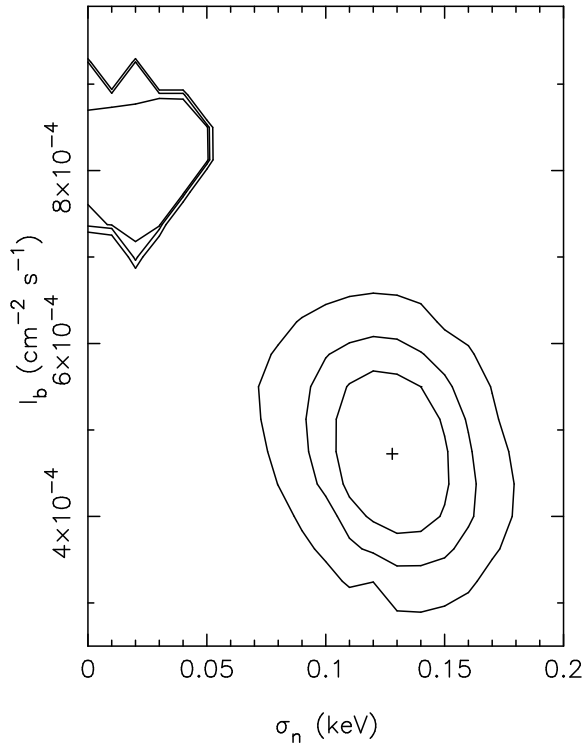}
\caption{The contour plots (1, 2, and $3\sigma$) of the photon flux in the disk
line, $I_{\rm b}$, vs.\ the width of the narrow line component, $\sigma_{\rm n}$
in the model with disk line only (W99), see \S \ref{w99}. There exist two
solutions with identical $\chi^2$, one with $\sigma_{\rm n}\sim 0$ and a strong
disk line, and one with $\sigma_{\rm n}\sim 0.1$ keV and the disk line twice as
weak. W99 have identified only the former.
\label{f:narrow}
}
\end{figure}

We have applied the same model to the SIS 1--10 keV data, and have closely
reproduced their results of $\Gamma\simeq 1.30$ and $\wfe \simeq 280$ eV at the
best fit (with $\chi^2/\nu =409/382$ with our binning and data reduction). At
this $\chi^2$ minimum, the width of the narrow line component is close to null.
However, we find there exists another minimum with the same $\chi^2$, at which
$\wfe \simeq 150$ eV and the width of the narrow component is $\sigma_{\rm
n}\approx 0.1$ keV, as illustrated in Figure \ref{f:narrow}. The uncertainties
on $\wfe$ given by W99, $\pm 20$ eV, appear to have been obtained under the
assumption of $\sigma_{\rm n}=0$. However, as we discuss in \S \ref{narrow}
below, the presence of strong variable absorption in \source\/ implies the
presence of Fe K emission of the absorbing clouds (Makishima 1986). If the
clouds are at a plausible distance of $\sim 10^3 GM/c^2$ and cover $\ga 10\%$ of
the $4\pi$ solid angle (see \S \ref{narrow}), $\sigma_{\rm in}$ will be just
$\sim 0.1$ keV. Thus, the data used by W99 do allow the disk line to be
substantially weaker than they claim.

On the other hand, the above ambiguity of spectral solutions disappears when the
GIS 1--10 keV data are added. In that case, there is only {\it one\/} $\chi^2$
minimum, which has $\wfe\simeq 150$ eV and $\sigma_{\rm n}\sim 0.1$ keV (similar
to other models with disk line presented in this work), and setting $\sigma_{\rm
n}=0.01$ keV results in $\Delta \chi^2=+21$. Furthermore, when we add the data
below 1 keV and model the soft excess in the same way as described in \S
\ref{simple}, we obtain $\chi^2/\nu =720/636$ (at $\Gamma=
1.47^{+0.06}_{-0.04}$, $\wfe = 110$ eV), which $\chi^2$ is {\it higher\/} by
$19$ for the same number of degrees of freedom as in the corresponding model
with reflection and dual absorber described in \S \ref{reflection}. Thus, the
reprocessed/reflected component associated with the Fe K lines is not only
needed physically but also {\it it is required statistically for our data}.

\section{Variability}
\label{variability}

\begin{figure}[t!]
\epsscale{1.0}
\plotone{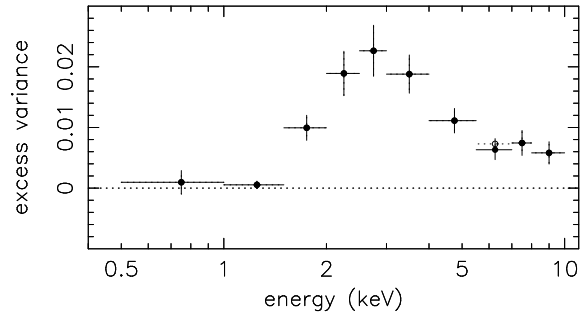}
\caption{The fractional amplitude of intrinsic variability squared
as a function of energy.  The dotted symbol in the 5.5--7.0 keV band
accounts for the constant narrow component of the emission line.
\label{f:var1}
}
\end{figure}

\begin{figure}[t!]
\epsscale{1.0}
\plotone{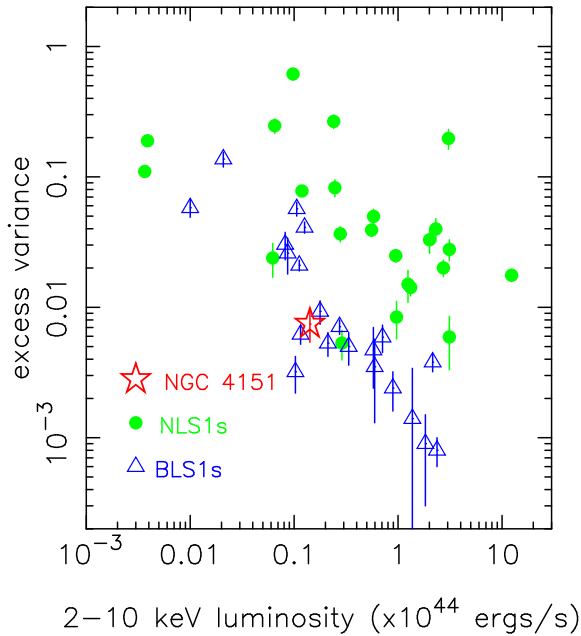}
\caption{Comparison of the intrinsic X-ray variance of \source\ with those of
other broad and narrow-line Seyfert 1s (based on Leighly 1999). The intrinsic
variance of \source\ is rather typical for other broad-line Seyferts. Note that
the decrease of the measured variance with increasing luminosity, $L$, appears
to be an effect of the decreasing ratio of the typical length of an observation
(limiting the range of timescales over which the variance is measured) to
$GM/c^3$ at $M\propto L$. The data shown here are compatible with the {\it
total\/} variance not showing any trend with $L$ within each of the two types of
Seyferts (e.g., Leighly 1999).
\label{f:var2}
}
\end{figure}

Overall, the total flux during the \asca\/ observation varied by $\sim 50\%$.
The complexity of the spectrum indicated above was born out in the considerable
spectral variability observed. The excess variance as a function of energy is
shown in Figure \ref{f:var1}. Below 1.5 keV the flux was constant, consistent
with emission from a extended region. We note that, given the energy dependence
of the variance shown in Figure \ref{f:var1}, the average \asca\/ variance of
the source is not a suitable quantity for comparison with other, much less
absorbed Seyferts (Nandra et al.\ 1997). On the other hand, the excess variance
in the 7--10 keV range, $7.5\times 10^{-3}$, is most likely to represent the
intrinsic source variability (although variability due to changes in the
absorption cannot be ruled out completely).  We have found this value to be
fully consistent with variance of other broad-line Seyferts having intrinsic
2--10 keV luminosity of $1.4\times 10^{43}$ erg s$^{-1}$ (e.g.\ Leighly 1999),
as illustrated in Figure \ref{f:var2}. This tends to rule out the Compton mirror
model proposed by Poutanen et al.\ (1996), which would predict a reduced
amplitude of variability unless the mirror has the same size scale as the
source.

Detailed analysis of the spectral variability is beyond the scope of this
paper. However, we note that the highest amplitude of variability is in the
2.5--3.0 keV band. This increase with respect to the 7--10 keV one (likely to
be mostly intrinsic) provides evidence for the importance of  variability in
the covering fraction/column density of the absorber in addition to the
spectral index/intrinsic flux variability. We have confirmed that our physical
model with complex absorption and disk line coupled to
relativistically-broadened reflection also fits the time-resolved spectra.
Thus, our fit results presented in \S 3 are not an artifact of time averaging.

The amplitude of variability is reduced in the 5.5--7.0 keV band compared with
neighboring energies, as shown in Figure \ref{f:var1}. This is most likely due
to the contribution of the constant narrow component of the iron line, which
contributes about 7\% of the flux in that band, implying that the excess
variance is a factor of 1.15 too low.  Accounting for that makes excess variance
in the 5.5--7.0 keV band comparable to that in the 7.0--8.0 keV band. The
constant narrow component supports our two-component (narrow and disk) model of
the iron line.  The constancy of the narrow component of the line in this data
set was first reported  by Leighly et al.\ (1998).

\section{The Narrow Line Component}
\label{narrow}

The flux of the narrow K$\alpha$ component, $I_{\rm n}\sim 2.5 \times 10^{-4}$
cm$^{-2}$ s$^{-1}$ in our best models (see Table 1; somewhat more in our other
models), is higher than the flux of $1.8\times 10^{-4}$ cm$^{-2}$ s$^{-1}$
observed by {\it Chandra\/} in 2000 March (Ogle et al.\ 2000).  Ogle et al.\
find that 65\% of the line (i.e., $\sim 1.2\times 10^{-4}$ cm$^{-2}$ s$^{-1}$)
comes from kpc-scale gas.  However, a part (dependent on the covering factor of
the absorber) of the narrow line has to come from the absorbing medium
(Makishima 1986), which medium is variable on long time scales (e.g., Yaqoob et
al.\ 1993).  We note that the optical flux peaked in 1995--1996, the U band
magnitude had decreased by $\Delta M_{\rm U} \approx 1.5$ by 2000 (Doroshenko et
al.\ 2001), and the {\it Chandra\/} observation was made when NGC 4151 was in a
low state.  We suggest that the flux of the narrow component was higher during
1995 than it was in 2000 (consistent with a long-term variability of the soft
excess, W94), and our results are consistent with those from {\it Chandra}. This
is, in fact, confirmed by the findings of Weaver et al.\ (2001), who obtained
$I_{\rm n}\sim (3$--$4) \times 10^{-4}$ cm$^{-2}$ s$^{-1}$ in their fits to five
\asca\/ observations of \source\ (in X-ray high states). Also, the recent {\it
XMM-Newton\/} result of Schurch et al.\ (2002), who found $I_{\rm n}\simeq
1.3\times 10^{-4}$ cm$^{-2}$ s$^{-1}$ in a very low state of \source\ (similar
to the historically lowest state shown in Fig.\ \ref{f:compare}) one year after
the {\it Chandra\/} observation supports the picture of a part of the narrow
line responding to changes of the intrinsic continuum.

The narrow component of the Fe K emission shows evidence of broadening, with
$\sigma_{\rm n}\simeq 0.05$--0.1 keV in our best models (see Table 1), and
somewhat more in disk line models with the standard dual absorber in \S
\ref{reflection}. This width is related to the velocity dispersion at the FWHM,
$\Delta v = (8\ln 2)^{1/2} (\sigma_{\rm n}/E_{{\rm K}\alpha}) c$ ($\la 10^4$ km
s$^{-1}$ at the above $\sigma_{\rm n}$). Roughly, $\Delta v$ will be similar to
the Keplerian velocity, which implies a characteristic radius where the emitting
clouds are located of $R_{\rm c} \sim 2\times 10^3 GM/c^2$. These clouds are
most likely the same clouds that absorb the intrinsic spectrum. As mentioned
above, {\it Chandra\/} and {\it XMM-Newton\/} measured much narrower Fe
K$\alpha$ lines at lower fluxes in later low states of \source\ (Ogle et al.\
2000). Thus, the ``narrow" Fe K$\alpha$ line appears itself to consist of two
components: one with $\sigma_{\rm n}\ll 0.1$ keV originating in an extended
region and constant on long time scales, and a component with $\sigma_{\rm
n}\sim 0.1$ keV emitted by the absorbing clouds and responding to variability of
the nuclear source. However, given the \asca\/ resolution, using one Gaussian to
represent both components of the narrow line is sufficient.

The characteristic line response time to changes in the nuclear continuum will
be $\sim 2R_{\rm c}/c$. For a fiducial $M=10^7 {\rm M}_\sun$, this gives $\sim
2\times 10^5$ s, consistent with the constancy (within the measurement errors)
of the narrow line component during our observation (Leighly et al.\ 1998; see
\S \ref{variability} above). On the other hand, variability of the absorber
during the observation ($\sim 10^5$ s) implies, assuming a cloud moving in the
line of sight with the Keplerian velocity, the characteristic cloud size of a
few times $10^{13}$ cm. This size should be similar to the size of the intrinsic
X-ray source in order to allow it to partially cover our line of sight. This, in
turn, is consistent with the size of the region where most of the gravitational
energy is dissipated, $\sim 20 GM/c^2$ or so.

The equivalent width of the narrow component with the parameters of Table 1 is
$\sim 60$ eV with respect to other components of the observed absorbed spectrum.
As calculated by Makishima (1986), the equivalent width from an absorber
completely surrounding a nuclear source would be $\sim 300$ eV at $N_{\rm H}\sim
3\times 10^{23}$ cm$^{-2}$ (as in Table 1). If a half of the observed $I_{\rm
n}$ comes from the absorber (and the rest from the extended gas), $\sim 10\%$ or
so of the lines of sight as seen from the nucleus should be covered by the
clouds.

\section{Discussion and Conclusions}
\label{discussion}

We have performed a detailed and careful analysis of contemporaneous long
observations of \source\ by the \asca\/ and \gro\/ satellites in 1995. We find
that proper inclusion of the Compton reflection component (taking into account
its own relativistic smearing) associated with the broad Fe K line in the source
is important for determination of the spectral index, $\Gamma$, of the power law
component in this source. This component is required physically as well it is
required statistically for our data (its inclusion results in a reduction of
$\chi^2$ by $-20$ without changing the number of free parameters). With it,
$\Gamma\ga 1.6$, which is within the $2\sigma$ range measured for Seyferts
(Nandra \& Pounds 1994). On the other hand, including Compton reflection has a
little effect on the fitted parameters of the broad line.

Furthermore, we have studied the complexity of the absorber in this source. As
previously noted by W94, the measured value of $\Gamma$ depends crucially on the
structure of the absorber. So far, the preferred model of the absorber has
consisted of a uniform neutral screen and a partially-covering cloud (the dual
absorber, e.g., W94). We have considered some simple and plausible modifications
to this model. First, we confirm a finding by Piro et al.\ (2002) that the
uniform screen is weakly ionized. Furthermore, we find very strong statistical
evidence ($\sim 1-4\times 10^{-6}$) for the presence of more than one
partial-covering cloud in the line of sight. We find that the presence of the
second absorbing cloud is not only strongly preferred statistically, but it also
changes the fitted value of $\Gamma$ to $\sim 1.9$, close to the average index
of Seyferts. In this model, the broad line component becomes relatively weak,
with $\wfe \sim 10^2$ eV (60--70 eV in our preferred physical model with thermal
Comptonization). The corresponding reflection strength is $\Omega/2\pi\sim 1$,
also average for Seyferts (Zdziarski et al.\ 1999). Our data also indicate that
the broad line component is dominated by the Fe K emission very close to the
central black hole, similarly to the case of MCG --6-30-15 (Wilms et al.\ 2001).

Obviously, our model is not unique and, e.g., further complexity of the absorber
is possible. One theoretical uncertainty involved regards the equivalent width
of the Fe K$\alpha$ line with respect to the reflection continuum. Our choice of
that quantity (1.3 keV) is based on George \& Fabian (1991), whereas \.Zycki \&
Czerny (1994) obtain values generally lower than 1 keV. Since our fits are
driven by the line strength, the above choice is conservative in the sense that
it gives the lowest theoretically predicted reflection strength. Adopting
results of \.Zycki \& Czerny (1994) would yield values of $\Omega$ and $\Gamma$
somewhat larger than those in Table 1.

Our model is consistent with a contemporaneous observation by the \gro/OSSE.
When the joint data are fitted by thermal Comptonization, the plasma temperature
is $\sim 70$ keV, in the range of typical values for Seyferts (Zdziarski et al.\
2000). We have also fitted data from \sax\/ observations of \source\/ and have
found excellent fits with the thermal-Compton model with two absorbing clouds
and Compton reflection (work in preparation).

The finding that \source\ appears to be an intrinsically average Seyfert 1 is
also confirmed by comparison of its X-ray variance with that of other
(optically) broad-line Seyferts. The crucial importance of complex absorption
for the appearance of the X-ray spectra is further supported by the energy
dependence of the variance, which peaks in the range where absorption of the
intrinsic power law is the dominant effect.

We also have studied the narrow component of the Fe K emission. We find the
presence of a broadened Gaussian component with $\sigma_{\rm n}\sim 0.1$ keV or
so in addition to the broad disk line. This allows us to construct a
self-consistent scenario (\S \ref{narrow}) with Fe K emission of the absorbing
clouds at a characteristic radius of a few thousands $GM/c^2$. This
moderately-broad component responds to changes of the nuclear continuum and thus
was nearly absent during {\it Chandra\/} observation of \source\ in a low state
(Ogle et al.\ 2000).

\acknowledgements

This research has been supported by grants from KBN (5P03D00821,
2P03C00619p1,2), the Foundation for Polish Science (AAZ) and NASA grant
NAG5-9745 (KML, AAZ).  Part of this work was done while KML was a STA fellow at
RIKEN.  We thank C. Done, T. Yaqoob, and the referee for valuable suggestions,
and W. N. Johnson and P.-O. Petrucci for the OSSE and \sax\/ data,
respectively.

\begin{deluxetable}{lccccccccccccccr}
\tabletypesize{\scriptsize}
\tablewidth{0pc}
\rotate
\tablecolumns{16}
\tablecaption{Rest-frame model parameters of the broad-band spectrum}
\tablehead{\colhead{} & \colhead{$N_0$} & \colhead{$\xi$} & \colhead{$N_1$} &
\colhead{$f_1$} & \colhead{$N_2$} & \colhead{$f_2$} & \colhead{$E_{\rm c}$,
$kT$} &
\colhead{$\Gamma$, $y$} & \colhead{$\Omega/ 2\pi$} & \colhead{$f_{\rm sc}$} &
\colhead{$kT_{\rm s}$} & \colhead{$E_{\rm n}$} &\colhead{$\sigma_{\rm n}$} &
\colhead{$I_{\rm n}$} & \colhead{$\chi^2/\nu$} }
\startdata
P &  $3.7^{+0.4}_{-0.5}$ & $0.37^{+0.27}_{-0.29}$ &$28^{+6}_{-6}$ &
$0.40^{+0.06}_{-0.08}$ & $6.0^{+0.7}_{-0.7}$ &
$0.77^{+0.05}_{-0.04}$ & $210^{+100}_{-60}$ &
$1.87^{+0.07}_{-0.08}$ & $0.95^{+0.29}_{-0.30}$ & 0.016 &
$0.16^{+0.02}_{-0.02}$ & $6.36^{+0.02}_{-0.02}$ & $0.07^{+0.02}_{-0.04}$ &
$2.4^{+0.3}_{-0.3}$ &682/658  \\
C & $3.8^{+0.3}_{-0.5}$ & $0.41^{+0.34}_{-0.28}$ &$32^{+5}_{-6}$ &
$0.44^{+0.05}_{-0.07}$ & $6.1^{+0.6}_{-0.6}$ &
$0.78^{+0.04}_{-0.05}$ &  $73^{+34}_{-29}$ &
$0.88^{+0.12}_{-0.11}$ & $0.60^{+0.24}_{-0.21}$ & 0.014 &
$0.16^{+0.02}_{-0.02}$ &  $6.35^{+0.02}_{-0.02}$ & $0.08^{+0.02}_{-0.03}$ &
$2.6^{+0.4}_{-0.2}$ & 682/658  \\
\enddata
\label{t:fits}
\tablecomments{Model P is a phenomenological model with an e-folded power law
while model C is a physical Comptonization model.
$N$, the ionization parameter $\xi$, $I_{\rm n}$, and ($kT$,
$E_{\rm c}$, $E_{\rm n}$, $\sigma_{\rm n}$) are in units of $10^{22}$
cm$^{-2}$, erg s$^{-1}$ cm, $10^{-4}$ cm$^{-2}$ s$^{-1}$, and keV,
respectively.}
\end{deluxetable}

\end{document}